Article

# A Proposal about the Meaning of Scale, Scope and Resolution in the Context of the Information Interpretation Process

**Gerardo L. Febres** [1,2]

[1] Departamento de Procesos y Sistemas, Universidad Simón Bolívar, Sartenejas, Baruta, Miranda, 1080, Venezuela; gerardofebres@usb.ve
[2] Laboratorio de Evolución, Universidad Simón Bolívar, Sartenejas, Baruta, Miranda, 1080, Venezuela.

**Abstract:** When considering perceptions, the observation scale and resolution are closely related properties. There is consensus on considering resolution as the density of the elementary pieces of information in a specified information space. On the other hand, with the concept of scale, several conceptions compete for a consistent meaning. Scale is typically regarded as a way to indicate the degree of detail in which an observation is performed. Surprisingly, there is not a unified definition of scale as a description's property. This paper offers a precise definition of scale and a method to quantify it as a property associated with the interpretation of a description. To complete the parameters needed to describe the perception of a description, the concepts of scope and resolution are also revealed with an exact meaning. A model describing a recursive process of interpretation, based on evolving steps of scale, scope and resolution, is introduced. The model relies on the conception of observation scale and its association to the selection of symbols. Five experiments illustrate the application of these concepts, showing that resolution, scale and scope integrate the set of properties to define any point of view from which an observation is performed and interpreted. The results obtained for descriptions expressed in one and two dimensions, are the basis for a comparison of the perceivable symbolic information from different interpretations of the same descriptions. In conclusion, this study provides a framework for building models of our interpretation process and suggests ways to understand some mechanisms in the formation of information from initially meaningless symbols.

## 1. Introduction

The 'anatomy' of a description has been the subject of intense discussion. Three abstract entities have been recognized as essential [1,2] for the construction of descriptions in any language or communication system: resolution, scale and scope. However, we have not found a unified definition of scale of a description. Scale is seldom treated as a quantifiable concept—one that can be managed by the computer—and the available definitions are rather ambiguous.

The concept of scale, along with its relationship to emergence and complexity, has been subject of research and discussion. Heylighen [1] presented emergence as a measure of the change of the dynamics after a system's transition. This measure cannot be directly taken from the system itself but from system models or observations of it, thus implying great relevance of the model's scale over the results of any measure. More recently, Bar-Yam [2,3] associated complexity with information profiles. Subsequently, Bar-Yam identified the relevance of the information that emerges when the system is observed from different levels of detail. In 2007, Ryan [4] depicted the relationship of scope, resolution and self-organization, considering emergence as the apparition of novel properties that a system exhibits when it changes from a condition to another. Ryan's discussion focuses on scope and resolution but the scale is left as a slave property of resolution. Later, in 2008, Prokopenko, Bochetti and Ryan [5] considered scale as a parameter defining emergence as a phenomenon. However, in their treatment, scale is conceived almost the same as the degree of detail or level of resolution, thus diminishing the independence that scale, as a concept, should have in relation to the scope and resolution.

Recently, Fernandez, Maldonado and Gershenson [6] indicated that any change of the system's structure reflects on the quantity of information required for its description. The change of the amount of information is a measure of the emergence process between any two system's statuses. Fernandez et al. [6] showed how four numbers, initially expressed in a sequence of binary digits, can be presented in a sequence of numbers expressed at different basis. The resulting entropy, computed for each message's string of characters, clearly suggests that there is an important impact of the base of the language used—the number of different symbols of the language—over the effort a reader must apply to interpret the message.

In the fields of spatial landscape representation and image analysis, several studies approach the effects of resolution over the delivered information. During the late 1980's, Turner et al. [7] presented the quality of a photograph as the result of the combination of resolution—which is sometimes called 'grain'—and scope—sometimes called 'extent'. Later, in 2003, Benz et al. [8] applied fuzzy analysis to recognize the relevant objects, which represent what we call symbols in this study. Relying on the comparison between the hypothetical object images and real-world-objects, they manage to build a "hierarchical object network" [8] where they somehow represent the issue of changing the observation-scale. Even though Benz et al. did not directly define scale as a property of the way a photograph is observed, they did recognize that the same photography may be given different interpretations. They also pointed out the relevance of these interpretations when looking for the best observation results.

In mathematics, a different concept of scale is used to adjust the summation of a series of wavelets to the value of a function. In 1909 Alfred Haar used a series of discrete periodic functions to represent target functions. During the 20th century the same idea of summing the terms of functional series to express the value of a function within a certain space, led to the now known as the Wavelet Transform. Zweig, Grossmann, Morlet, Strömberg, Daubechies, Meyer, Mallat and Zhong are some researchers involved in the development of the Wavelet Transform and the Scale-space theory; a field devoted to the representation and processing of signals. In this field, the scale represents a parameter used to define the "degree of smoothing" applied to the processed signals. This processing discards fine details as well as coarser information, thus acting as a filter which results useful for information-compression purposes and mathematical modeling.

Enhancing the observation of a system's description has been typically restricted to the grouping of information-elements into spaces topologically congruent. This approach has proved to be of limited utility because it restricts the effects of varying parameters within the process of interpreting descriptions. Complex systems offer difficulty to the recognition of all their components. Depending on the focus of the observer, some parts of the system may be recognized as elements while other parts may be shadowed. Additionally, the frontier separating the perceived elements may not be clearly defined due to overlapping, or on the contrary, due to empty or meaningless sectors, which do not add information to the perceived system's description. Whatever the situation is, the observer decides which are the elements forming his or her interpretation. When the observer associates some sectors of the description with some meaning or logical pattern, he or she decides which are symbols representing the elements forming his interpretation of the observed description. Thus, the selection of the symbols which integrate the description is a crucial step in the interpretation process. Considering the number of symbols and their size is closely related with the concept of scale. Thus, there seems to be a link between the meaning of scale and the set of symbols participating in the interpretation of a descriptive process.

This paper offers a novel conception of scale, pointing out clear differences with the concepts of resolution and scope. Several experiments using descriptions of diverse nature are used to apply our conception of scale, resolution and scope. The impact of interpreting these descriptions at several scales is evaluated for each of these experiments. Among the possible ways of looking at a system's description, there is one we give special attention to: The Fundamental Scale. Initially introduced by Febres and Jaffe [9], the Fundamental Scale is the scale at which an observation can be interpreted with minimal entropy. Some formalization of the concept of Fundamental Scale, is also an objective of this paper.

This study is organized as follows:

- Sections 2 and 3. Proposal about scale, scope and resolution. These sections present definitions and intuitive notions of scale, resolution and scope, valid for multidimensional descriptions. The concept of the alphabet, as a set of elementary symbols, is also included in these sections.
- Section 4. The interpretation process is hypothesized over the selection of symbols with the criterion of minimizing the description's entropy, computed over a written version of the message.
- Section 5. The proposed concepts of scale, resolution and scope are used to evaluate descriptions of different nature. Five experiments are performed to evaluate the information content of each experiment's description seen at different scales of observation.
- Sections 6 and 7. Discussion and conclusions about the implications of the proposed meaning of scale and the interpretation process are presented.

## 2. Resolution, Scale, Scope and Other Properties of Descriptions

### 2.1. Resolution

Resolution is a human created artifact to split system descriptions in regular, equally shaped and sized, pieces. It results from the process of discretizing the description of a system. The original description can be discrete or can be an analogous depiction taken directly from physical reality. Resolution is typically regarded as the number of equally sized pieces in which we divide the original description and thus it refers to the size of the smallest piece of information of a description, or equivalently, to the density of elementary-pieces of information. The resolution is commonly specified as the number of smallest information pieces that fit into each dimension of the description. Considering all dimensions conforming the description, the resolution can be regarded as the total number of elementary information pieces included in the whole description. In this case we use the letter $R$ to refer to it. When resolution is specified as the density of information contained in a physical dimension, we use the number information pieces $r_j$ that fit into the dimension considered, thus $r_j = R_j/Dim_j$, where $Dim_j$ refers to the absolute size of the physical dimension used. As an example, we can consider a 16″ × 9″ computer screen with 1920 pixels in the horizontal longitudinal dimension and 1080 pixels in the vertical longitudinal dimension. The resolution $R$ is regarded as 1920 × 1080 (pixels × pixels) and $r$ would be 120 × 120 (pixels/in × pixels/in). In another situation, if the description refers to a 60-seconds long sequence of 36,000 very short sounds, then $R = 36,000$ (sounds) and $r = 600$ (sounds/s).

Resolution, as a concept, loses meaning when the mesh of information elements is not regular as it would be an information structure formed by a set of symbols with a diversity of sizes and shapes. Such a situation can hardly be described using the resolution as a characteristic parameter because the density of the resolution would not be a constant.

### 2.2. Scale

All of us have an intuitive notion of scale. Commonly, the term scale is associated with the distance from which the system is observed. Thus, the term 'scale' is typically used to mean that the system is being interpreted from a closer point of view—higher scale with finer detail, or from a further point of view—lower scale with less detail. This works fine if the scale is exclusively established by the distance between the observer and the object but this not always the case. In fact, the size of an element within the perception of a system, does not necessarily represent the relevance of that element as part of the system's description. Thus, the observation distance cannot reflect in a sole parameter the property defined as 'scale'. Additionally, the frequent exchange of the terms 'detail of degree', 'level', 'observation distance' and 'scale', promotes the confusion between the concepts of scale and resolution. I think this confusion of terms has hindered a clear conception of what a scale actually is and as a result, we are still lacking a unified definition of scale of a description that can work in an information space with non-regular shaped symbols.

During the last decade, several studies reflected the relevance of the concept of scale in our interpretation of descriptions. In 2004, Bar-Yam [2,3] presented complexity as a property intimately related to the scale. His treatment of scale as a parameter capable of varying continuously rather than discreetly, led him to present the so called 'scale profiles'; a sort of 'structure' relating entropy to scale, where entropy decreases monotonically as the observation scale shows less detail. Even though these profiles are a good representation of the envelope of the complexity, they would only be constructible by representing scale as the result of the increase, or decrease, of the resolution with which the system is observed. Thus, these profiles illustrate the relationship between complexity and resolution, while the actual scale parameter is left out of the analysis.

Piasecki and Plastino [10] presented entropy as a function of scale length. They defined scale as the size of the group of pixels used to represent each object in a pattern of regularly distributed grayscale pixels. Their graphs show local minimal values of entropy when the scale length is a multiple of the characteristic size of the pattern measured in pixels. Again, Piasecki and Plastino [10] actually work with resolution changes but their experiment shows how, when the size of the objects described fits with an integer number of pixels, the entropy drops, implying that the arbitrary choice of the number of pixels used to describe an object modifies the entropy, thus the system's observed complexity. This confirms that the observation scale is not either exclusively dominated by the system properties nor the resolution.

Scale seems to be a result of our process of interpretation. When we consider a description, our brain probably scans several interpretations of the observed description. At each interpretation, we combine raw information by joining adjacent information elements and forming with them larger hypothetical symbols. Simultaneously, we look for patterns which we can associate with previous experiences and learned notions, or even with our personal conceptions of beauty, thus giving certain meaning to a message that was initially abstract. Therefore, the scale is a property of the way the observer looks at the system's description. Once the observed system's conception is 'organized' in our brain, a clear account of the symbols resulting from our interpretation, along with their frequency of appearance and their relative position, constitute our model of the description. This lets me introduce statements about the scale that do not contradict our previous intuitive notions: the scale of a description, as it is observed, is directly related to the set of symbols used to create the description's model. The scale can be numerically represented by the number of different symbols used in each interpretation, thus the numerical value of the scale works as the base of the language we use for our interpretation. The numerical system we use, for example, is regarded as a base = 10 system because it consists of 10 different digits. If we use only two symbols, then we are reading the system at base = 2. The physical system itself has not to be binary but we are interpreting it by means of a binary language.

When symbols fit into a regular lattice of pixels, the number of pixels forming each one of these regular symbols, specify the shape and the size of them. If this were the case, saying the system exhibits rectangles formed by $n \times m$ pixels could be appropriate to specify the manner we are looking at the system's description, i.e., to specify our scale of observation. If on the contrary, we are looking at some irregular-shaped objects such as the countries on a map, there are no constant arrays of pixels that can be assigned to indicate that we are looking at countries. In this case the symbols should be the countries shown in the map, disregarding any number of pixels contained in any country; the base of the scale would be the number of different countries seen in the map. If we see the same map at the scale of continents, the symbols become the continents and the scale's base would be the number of them.

Finally, it should be emphasized the quantitative notion of scale, or the base of the scale, which value is identical to the symbolic diversity and therefore the designation of *D* is commonly used to refer to both, diversity and scale's base, or simply scale.

*2.3. Scope*

The scope refers to the total number of information-units contained in the description. When an n-dimensional description is performed over arrays of regularly distributed elements—or symbol, the scope equals the resolution of each symbol times the *n* dimensions occupied by the description. Up to this point, the scope seems to be a redundant concept with resolution. However, when the symbols are contained—or organized—in a non-regular sized mesh, the resolution becomes a more difficult parameter to represent in numerical terms; nevertheless, we could still use the scope to characterize the description, just by counting the number of information elements, or symbols, contained.

Relaxing the condition of interpreting the description over a regular mesh allows for looking at different sized and shaped symbols to integrate the complete description. Once the size and shape of the symbols have been established, the entire description conveys an amount of information determined by the total number of symbols, repeated or not, included in the description. In this sense, the scope equals the length of the description measured as the number of symbols and thus, length and scope are both represented by the letter *L*.

*2.4. Symbols, Alphabet and Encoding*

The alphabet is only indirectly connected to a description. An exact definition of our conception of the alphabet is justifiable to properly depict some notions about scale. The alphabet is the set of elementary symbols used to create the compound symbols that may be part of our interpretation of a message. In other words, a symbol may be formed by several alphabetic symbols—usually regarded as characters. When reading an English text on a computer screen, the symbols of the description could be the words of the text, while the alphabet should be considered as the set of letters used to form larger symbols or words. While these larger symbols are useful due to their capacity to store meaning or semantic content, the symbols forming the alphabet are tools with no meaning by themselves. Even though alphabet symbols can be represented as the conjunction of even smaller symbols, at a certain point of our interpretation, we are not interested in the sub-symbols of the alphabet symbols. Thus, when talking about a natural language, for example, the characters and not the pixels, are considered as the elementary symbols which form the corresponding alphabet.

The symbols of an alphabet are useful to create larger and probably more meaningful, symbols. We refer to the rules which allow the formation of alphabet symbols as the 'encoding'. With this schema, an English text can be regarded as a description formed by words, which in turn are formed by letters. These letters are elements of the set considered as the English alphabet. Now, each one of these letters may be written by a single drawing, as with a pencil and paper, or by another 'letter-forming system' as a group of pixels with color and shape resembling the alphabetic symbol or, if the encoding Morse system is used, the set of dashes and dots. This reasoning leads to treating the alphabet and encoding as artifacts, which enables us to construct an observation scale. However, the alphabet and the encoding do not directly link to the description, though they are essential to any written communication system. Whenever the alphabet is confused with the most relevant set of symbols used in a description, the freedom to form effective interpretations of the message, disappears. In this regard, exploring the impact the alphabet has over the effectiveness of a language seems to be an interesting branch of research. However, it lies beyond the scope of the present study.

## 3. More about the Meaning of Scale

*3.1. An Intuitive Notion of Scale*

The scale of a description has been commonly considered as groups of the most elemental information components used within the description. If, for example, the description consists of an image projected on a computer screen, the pixels would be the most elemental components because each pixel shows the same color and therefore it would not make sense to divide it into smaller pieces. Assume now the screen shows letters one after another forming an array of characters that fills up the screen. Then, each letter is formed due to the contrast between the group of adjacent pixels and the surrounding pixels but when focusing our attention to each letter, we disregard the elementary pixels and interpret the shape of the whole letters as symbols. Then we can say that our observation is on the character-scale.

*3.2. The Alphabet as a 'Coding' Tool*

People speak, read and write natural languages using words as symbols to pack any meaning they want to incorporate to a more complex entity: the idea. When communication occurs by means of patterns of sonic, graphic, chemical signals or any physical media, the meaningful ideas or instinctive triggers are associated with these patterns. No transcription is needed because each receiver already has an interpretation and meaning about the signal pattern received. It is worthwhile to mention that in the absence of meaning—any meaning including abstract meaning—implies considering the involved set of symbols as noise.

The written version of a message is just the coded version of the idea the sender intends to communicate. Here, we consider an alphabet as the tool that offers the elementary symbolic pieces to allow for the coding of a message and to 'write' it onto a surface or any registering media. Thus, for example, for western natural languages the alphabet is the set of letters we use to record in a physical medium the syllables which in turn represent simple sounds. For Asian languages, it works differently; the alphabet is formed by a large collection of symbols, each one representing a sound with a meaning in its corresponding language and therefore there is no need to form multi-character written syllables. For music, the alphabet can be considered the set of figures used to represent a sound with a pitch specified by the position of the figure over the pentagram. The music figure's shape itself indicates the lasting of the sound.

This is important because we analyze information by handling some sort of written version of any message. In terms of analyzing and figuring information, we do not know how to directly deal with the perceptions of our senses. Therefore, to operate and compute properties of a description expressed in a specific language, we do not use the original version of the message. Instead, we rely on a written version that allows us to split and group different parts of the message and look for logical patterns which may represent information.

The writing systems then, are crucial for our method because we are actually evaluating the properties of communication systems by means of their corresponding writing systems. To illustrate this point, the Table 1 shows a non-exhaustive list of several languages with some associated writing systems and their alphabets. The proposal is then, to recognize those sequences of characters which can be used as information elementary units to form a message transcription.

**Table 1.** Examples of Languages of different nature and their corresponding writing and encoding systems.

| Nature/Dimensions | Language Group | Examples | Alphabet Components | Encodings |
|---|---|---|---|---|
| Natural Languages 1 Dimension | Alphabetic | English Spanish Russian French Arabic | letters Punctuation signs | Conventional writing Binary Braille Morse Code ASCII code |
| | Syllabic | Chinese Korean Japanese | Phonograms | |
| Sonic 1 Dimension | Music | Western music Indu-raga music Chinese music | Chromatic scale Major scale, Minor scales Pentatonic scale Other Scales Sound Volume Alteration signs Rhythms Tempos | Pentagram Notation Cypher |
| | | African Traditional Music | Rhythms, Tempos | |
| | Alarm and Warning signals | Alarms Ringing tones Warning Calls | Stridency of harmonies Ascending or descending tonal patterns | Characters of a recording Amplitude and Frequency of analog recording |
| Graphic 2 Dimensions | 2D graphical expressions | Painting | Color spectrum Brush Strokes | Pixel Size Pixel Color |
| | | Charts & Diagrams | Geometric Shapes | Shapes, colors, borders |
| Mathematics 1+ Dimensions | Numerical system, 1 dimension | Numbers | Digits and decimal point | 0, 1, 2, 3, 4, 5, 6, 7, 8, 9, '.' Binary, any other base |
| | Equations, 1+ dimensions | Math Expressions | Mathematical operators Char-strings declared as Symbols | Conventional Math ASCII code Binary |
| Biology 1 Dimension | Genomes | ADN | Adenine, Thymine, Cytosine, Guanine | A, T, C, G |
| | | RNA | Adenine, Uracil, Cytosine, Guanine | A, U, C, G |

## 4. The Relationship between the Interpretation Process and Scale

This section is devoted to describing some of the mechanisms driving the interpretation process. First, we present a model to quantify the information obtained from different interpretations of the same message. Three easy-to-follow examples with a string of characters are included. Afterwards, three possibilities for symbol-evolution pursuing entropy reduction are explained. The section ends presenting the interpretation process and some additional concepts involved in it.

*4.1. A Model to Quantify the Information of Different Interpretations*

Consider a description $\mathcal{M}$ as the sequence of symbols $A_i$ all of them contained in the alphabet $\mathcal{A} = \{A_1, A_2, A_3, \ldots, A_n\}$. Thus, $\mathcal{M}$ can be seen as a specific sequence of the letters (or characters) of alphabet $\mathcal{A}$. A descriptor of $\mathcal{M}$ is the probability distribution $P(A_i)$ where $A$ is the random variable associated with the frequency with which each letter of alphabet $\mathcal{A}$ is encountered within message $\mathcal{M}$. Then, using $f_{A_i}$ to represent the number of times the letter $A_i$ appears within the message, we can model message $\mathcal{M}$ writing:

$$\mathcal{M}: P(A) = \{P(A = A_1), P(A = A_2), \ldots, P(A = A_i), \ldots, P(A = A_n)\} = \{f_{A_1}, f_{A_2}, \ldots, f_{A_n}\}. \tag{1}$$

Then, the entropy $h_{\mathcal{A}n}$ associated to the message $\mathcal{M}$ when observed as the $n$ different letters of alphabet $\mathcal{A}$, equals the entropy of distribution $P(A)$,

$$h_{\mathcal{A}n} = -\sum_{i=1}^{n} f_{A_i} \log_n f_{A_i}. \tag{2}$$

The figure computed in Expression (2) holds for the entropy obtained when reading the message one letter at a time; each letter would be considered as a symbol and no symbol would be allowed to be formed by more than a single letter. This is what is called '*reading at the scale of letters*'. We know this would be a very ineffective way of reading. Fortunately, there are better ways of reading message $\mathcal{M}$ by grouping neighbor characters $A_i$ to form $D$ different multi-character symbols $Y_j$. This approach may be effective in the sense of reducing the associated entropy. Even more, assume the symbols $Y_j$ are such that the resulting entropy is minimal, we can write:

$$h^*_{\mathcal{A}D} = -\sum_{j=1}^{D} P(Y_j) \log_D P(Y_j), \tag{3}$$

where $h^*_{\mathcal{A}D}$ represents the minimum possible entropy associated to the process of interpreting the message encoded by means of alphabet $\mathcal{A}$. In order to comply with the character neighborhood condition when forming symbols $Y$, any symbol $Y_j$ must comply with

$$Y_j = A_i \, A_{i+1} \, A_{i+2} \ldots A_{i+S-1}, \tag{4}$$

where the sub-index $S$ represents the length of symbol $Y_j$. The set of symbols $Y$ is useful to read and interpret message $\mathcal{M}$, then we consider it a language. Since using language $Y$ is a better way of reading $\mathcal{M}$, we replace the messages formerly expressed in (1) model with:

$$\mathcal{M}: P(Y) = \{P(Y = Y_1), P(Y = Y_2), \ldots, P(Y = Y_j), \ldots, P(Y = Y_D)\}. \tag{5}$$

Notice that now symbols $Y_j$ are of different sizes. Therefore, the probability $P()$ of encountering a symbol $Y_j$ within the message, should account for the frequency of appearances $f_{Y_j}$, the size of the symbol $S_{Y_j}$ and the length $L_{\mathcal{A}}$ of the message measured in characters. Thus, this probability can be expressed in terms of the frequency $f_{Y_j}$ and the number of times the symbol of size $S_{Y_j}$ fits into a $L_{\mathcal{A}}$ characters long message, as

$$P(Y_j) = \frac{f_{Y_j} \cdot S_{Y_j}}{L_{\mathcal{A}}}. \tag{6}$$

Equation (3) for minimal entropy can now be rewritten in terms of the symbols of language $Y$ built over the characters of alphabet $\mathcal{A}$. We obtain:

$$h^*_{\mathcal{A}D} = -\sum_{j=1}^{D} \frac{f_{Y_j} \cdot S_{Y_j}}{L_{\mathcal{A}}} \log_D \frac{f_{Y_j} \cdot S_{Y_j}}{L_{\mathcal{A}}}. \tag{7}$$

To illustrate these figures, we present examples consisting of different observations of the same description and their entropy evaluation. Consider the following character-sequence as the description, where in order to ease the reading, the character 'Ø' appears in place of the 'space-character':

$$Øa\ ØabØabc\ Øabcd\ Øabcde\ Øabcdef\ Øabcdefg$$

Using $Y$ as the random variable representing the probability of encountering a specific symbol within the description, three examples are shown below:

**Example 1 (E1).** *Focusing on single characters as the symbols of this description, leads to the following probabilities based on their frequency:*

$$P(Y_1 = 'Ø') = \frac{7}{35}, \quad P(Y_2 = 'a') = \frac{7}{35}, \quad P(Y_3 = 'b') = \frac{6}{35}, \quad P(Y_4 = 'c') = \frac{5}{35},$$

$$P(Y_5 = 'd') = \frac{4}{35}, \quad P(Y_6 = 'e') = \frac{3}{35}, \quad P(Y_7 = 'f') = \frac{2}{35}, \quad P(Y_8 = 'g') = \frac{1}{35}.$$

*Since there are eight different symbols, the scale base is* $D = 8$ *and the computed entropy, applying Equation (3), is* $h \approx 0.937$.

**Example 2 (E2).** *A second interpretation of the same description assumes the message was expressed with a natural language and therefore regards as symbols those character strings which look as words. Those are the symbol strings beginning or ending with a 'space' ('Ø'). Under these conditions the symbols of the description would lead to the following probabilities:*

$$P(Y_1 = 'Øa') = \frac{1 \cdot 2}{35}, \ P(Y_2 = 'Øab') = \frac{1 \cdot 3}{35}, \ P(Y_3 = 'Øabc') = \frac{1 \cdot 4}{35}, \ P(Y_4 = 'Øabcd') = \frac{1 \cdot 5}{35},$$

$$P(Y_5 = 'Øabcde') = \frac{1 \cdot 6}{35}, \ P(Y_6 = 'Øabcdef') = \frac{1 \cdot 7}{35}, \ P(Y_7 = 'Øabcdefg') = \frac{1 \cdot 8}{35}.$$

*The scale base is* $D = 7$ *and the computed entropy is* $h \approx 0.957$.

**Example 3 (E3).** *Another interpretation looks for sequences of characters to form symbols that leads to a reduction of the computed entropy. The following selected symbols depict the message while their probabilities and sizes reduce the message's entropy to its minimal value:*

$$P(Y_1 = 'Øabc') = \frac{5 \cdot 4}{35}, \ P(Y_2 = 'de') = \frac{2 \cdot 2}{35}, \ P(Y_3 = 'ØaØa') = \frac{1 \cdot 4}{35}, P(Y_4 = 'b') = \frac{1 \cdot 1}{35},$$

$$P(Y_5 = 'd') = \frac{1 \cdot 1}{35}, P(Y_6 = 'def') = \frac{1 \cdot 3}{35}, P(Y_7 = 'f') = \frac{1 \cdot 1}{35}, P(Y_8 = 'g') = \frac{1 \cdot 1}{35}.$$

*The scale base is* $D = 8$ *and the computed entropy is* $h \approx 0.689$.

*4.2. The Entropy Reduction Process*

A way to study the change of the resulting entropy, produced by the change of the selected set of symbols, is by focusing on one or two consecutive symbols. For a single symbol the only way it can change itself while keeping invariant its character sequence, is by splitting itself into two or more symbols. For two adjacent symbols, we identify two cases in which they can vary while keeping constant the text represented by the two symbols together: symbols' boundary drifting and symbols joining. Examples of these cases are presented.

***Case 1. Symbol Splitting:*** Split symbol $Y_j$ into two smaller symbols. In this case, the object symbol $Y_j$ is replaced by new symbols $Y'_j$ and $Y'_{j+1}$. The sub-indexes of all successive symbols will increase by one. The new symbolic diversity also increases from $D$ to $D + 2$. One could think that symbol $Y_j$ as the only instance of that symbol in the entire text. If that were the case, by replacing it with $Y'_j$ and $Y'_{j+1}$, it would generate only one additional symbol. However, this seems very unlikely because, in order to be part of a minimal entropy set of symbols, $Y_j$ should appear more than once in the message. Another possibility is that the new symbols $Y'_j$ and $Y'_{j+1}$ already exist within the message. This is also unlikely because the minimal entropy criterion has already '*decided*' to represent the sequence $Y'_j Y'_{j+1}$ with the '*compacted*' form $Y_j$. Concluding, for Case 1 we can use $V = D + 2$ as the symbolic diversity for this message interpretation. Example 4 (E4) illustrates this situation:

**Example 4 (E4):** *Take* $Y_3 = \{\text{\O}abc\}$ *from the interpretation of Example 2. Then, symbol* $Y_3$, *could be split into two symbols called them* $Y_{31}$ *and* $Y_{32}$. *Making* $Y_{31} = \{\text{\O}ab\}$ *and* $Y_{32} = \{c\}$ *is a valid replacement for* $Y_3$ *since they add up the original* $Y_3$ *with no character overlapping. In this case the emergence of symbols* $Y_{31} = \{\text{\O}ab\}$ *and* $Y_{32} = \{c\}$, *which did not appear in the previous interpretation, would increase the symbolic diversity by 2 and since* $Y_3$ *only appeared in one instance, now extinguished, the resulting diversity after the symbol splitting end up being* $D$ *to* $V = D + 1$. *The resulting symbol set probabilities are:*

$$P(Y_1 = '\text{\O}a') = \frac{1 \cdot 2}{35}, \ P(Y_2 = '\text{\O}ab') = \frac{1 \cdot 3}{35}, \ P(Y_3 = '\text{\O}abc') = \frac{0 \cdot 4}{35} = 0, \ P(Y_4 = '\text{\O}abcd') = \frac{1 \cdot 5}{35},$$

$$P(Y_5 = '\text{\O}abcde') = \frac{1 \cdot 6}{35}, \quad P(Y_6 = '\text{\O}abcdef') = \frac{1 \cdot 7}{35}, \quad P(Y_7 = '\text{\O}abcdefg') = \frac{1 \cdot 8}{35},$$

$$P(Y_{31} = '\text{\O}ab') = \frac{1 \cdot 3}{35}, \ P(Y_{32} = 'c') = \frac{1 \cdot 1}{35}.$$

*The resulting scale base is* $D = 8$ *and the computed entropy is* $h \approx 0.926$.

*Case 2. Symbols Boundary Drifting*: We refer to 'boundary drifting' as the effect produced when the end and start of neighbor symbols shifts, leaving constant the total length of both symbols but adding to a symbol the same number of characters diminished on the other. The boundary between two adjacent symbols $Y_j$ and $Y_{j+1}$ moves making one symbol larger and the other smaller and resulting in new symbols $Y'_j$ and $Y'_{j+1}$. The sub-indexes of symbols do not shift but the new symbolic diversity increases from $D$ to $D + 2$ due to the *birth* of new symbols $Y'_j$ and $Y'_{j+1}$ which did not exist before. Again, as reasoned for Case 1, if the symbols $Y'_j$ and $Y'_{j+1}$ had existed before, they would likely appear as minimal entropy symbols $Y_j$ and $Y_{j+1}$, prior to the boundary drifting. Thus, for Case 2 we can use $V = D + 2$ as the symbolic diversity for this message interpretation.

**Example 5 (E5):** *Take* $Y_1 = \{\text{\O}abc\}$ *and* $Y_2 = \{de\}$ *from the interpretation of the Example (E3). Then we see the sequence* $Y_1 Y_2$ *('\text{\O}abcde') as the sequence of two new symbols* $Y_{11} = \{'\text{\O}ab'\}$ *and* $Y_{21} = \{'cde'\}$, *leaving the original message intact. The emergence of symbols* $Y_{11} = \{\text{\O}ab\}$ *and* $Y_{21} = \{cd\}$ *may compensate for frequency reduction of symbols* $Y_1 = \{\text{\O}abc\}$ *and* $Y_2 = \{de\}$, *which may extinct after the replacement of these instances. In any case, after each symbol boundary drifting step, the resulting symbol diversity will remain the same or increase at most from* $D$ *to* $V = D + 2$. *In this Example, the resulting symbol set probabilities are:*

$$P(Y_1 = '\text{\O}abc') = \frac{4 \cdot 4}{35}, \ P(Y_2 = 'de') = \frac{1 \cdot 2}{35}, \ P(Y_3 = '\text{\O}a\text{\O}a') = \frac{1 \cdot 4}{35}, P(Y_4 = 'b') = \frac{1 \cdot 1}{35},$$

$$P(Y_4 = 'd') = \frac{1 \cdot 1}{35}, P(Y_5 = 'def') = \frac{1 \cdot 3}{35}, P(Y_6 = 'f') = \frac{1 \cdot 1}{35}, P(Y_7 = 'g') = \frac{1 \cdot 1}{35},$$

$$P(Y_{11} = '\text{\O}ab') = \frac{1 \cdot 3}{35}, P(Y_{21} = 'cde') = \frac{1 \cdot 3}{35}.$$

*The resulting scale base is* $D = 10$ *and the computed entropy is* $h \approx 0.785$.

*Case 3. Symbols Joining*: Two adjacent symbols $Y_j$ and $Y_{j+1}$ are replaced by one larger symbol $Y'_j$. Depending on the number of instances of symbols $Y_j$ and $Y_{j+1}$, the symbolic diversity may reduce by one or two symbols, or may not reduce at all. Thus, the new diversity will be in the interval $V = [D - 2, D]$.

Starting from the fundamental set of symbols, there are many other ways of modifying the selection of symbols but all of them can be understood as the result of combining Cases 1, 2 and 3.

*4.3. The Interpretation Process*

Based on the schema presented above, the interpretation process can be understood as a repetitive process of symbol selection. Starting from an arbitrary symbol-set to integrate the message, the process of interpretation steps over the splitting, joining and the boundary-drifting of symbols, pursuing a reduction of the entropy associated to the set of selected symbols. Describing these steps as a process directed by the selection of the symbol-set to achieve a reduction of entropy, may sound unreal. We may not be conscious about the resulting entropy of one or another symbol selection. However, when observing at a description, for instance a picture, one of the first things we do is to locate the limits of the objects within the picture. Whenever these limits are not easily recognized, we experience difficulty to understand the description in front of us, then we strive for encountering a contour or a shape which helps us to find a way in the interpretation of the whole picture. This process continues until our interpretation settles in the states where we are able assign an intelligible meaning to the whole message, until we are satisfied with the sensations triggered by the message, or until we reject paying more attention to it because only noise was found. When there are obvious, meaningful patterns, our interpretation quickly fixes those patterns and adjusts the boundaries and sizes of the surrounding symbols searching for a coherent contextual meaning within the description's scope. This is the case of written or spoken words in natural languages. When we can communicate with a natural language, we rapidly recognize the symbols of a description; thus, in this case the interpretation process is not as complex as it may appear. As we are knowledgeable about the language, we can even detect—and correct or compensate for—deviations that may exist with respect to the accepted forms of the language. We can imagine this 'acceleration' of the interpretation process, not only for natural languages but also for other radically different means of transmitting information. Take, for example, the case in which we see a pattern of bands of seven different colors: red, orange, yellow, green, blue and violet. In that order. For most of us that pattern has an already assigned semantic meaning; in spite of its apparent arbitrary and abstract condition, whether we read it or see it, this pattern refers to a rainbow. If the pattern of colors is shaped as a bow, the association to a rainbow is intensified. Sound signals may also form patterns of symbols we can interpret according to a similar process, just working with sonic perceptions instead of colors or graphical patterns. Of course, this is an extremely simple and raw explanation of the interpretation process taking place in our brains, yet, it may be a basis to extend our understanding about perceptions and interpretations.

Only symbolic information is considered here for purposes of quantitative calculations. Semantic information, as the form of information we use to store most of our experiences and learning, has been left aside. However, using this hypothesis about the mechanism of symbol selection, we have built the Fundamental Scale Algorithm (FSA) [9] which controls a computer searching for the most relevant symbols included in an unintelligible sequence of characters and to extract from it, nearly the maximum possible information. Although the FSA is now developed for one-dimensional descriptions, these results strongly suggest that languages evolve to develop their capacity to organize symbols to effectively convey information. In this evolution process, the symbols adopt sizes and shapes that drive the shape of the resulting symbol-resolution groups away from regular patterns.

*4.4. The Scale Is a Choice*

The observation scale does not have to be based on a regular lattice. When nature 'decides' how to group its elements to constitute organisms and societies, it rarely pays attention to the regularity of the spaces that serve as frames for those compound entities. Of course, there are cases that can be seen as exceptions to this statement. The hexagonal pattern in a panel of bees or the fractal describing the appearance of a fern, could be considered regular spaces but in a description combining panels of bees and ferns, the resulting space would be hardly considered regular. Thus, when observing the elementary information pieces of a description, for example, letters in an English-written text, the question is how to group those letters in order the get the *best* possible understanding of the text; for a natural language we already know, the answer would probably be: words.

In the communication process, there is always present pressure to make the language used, more effective. As Zipf [11] stated in his "Principle of the Least Effort", we tend to reserve shorter words to apply them for the more frequent ideas we need to express. Following Kolmogorov's "Shortest Description Length", in 1978 Rissanen [12] introduced the Minimal Description Length Principle (MDL. Rissanen used the name "Shortest Data Description") to build minimal description-length models of an observed sequence of characters. The MDL has been widely studied afterwards (Hansen and Yu [13]). Most studies are centered on the problem of the compressibility of strings and approach the MDL problem as the construction of the probability distribution model which best resembles the real probability distributions of symbols beneath a description.

The typical approach considers the binary alphabet as the base for all descriptions, thus considering the binary alphabet as the base for the descriptions whose length is measured and is eventually compressed. In the Computer Science field this is justifiable because in digital computers all files, strings, models, etc., are at some point stored and handled through binary codes. Nevertheless, we are concerned with the language itself and therefore, this proposal is about the search of the alphabet in which the description was naturally written. Doing so, we highlight the fact that any language has the tendency to evolve around its adaptation in order to produce short descriptions.

A hypothesis is then, that for every language, there is a set of symbols which reduce at a minimum the lengths of the descriptions built with it, thus minimizing their entropy based on the frequency of the symbols. Along with their relative frequencies, this set of symbols constitutes a good model of the language and represents what the language itself has chosen as its underlying structure; it is an appropriated scale to use when interpreting a description. Since this hypothetical set of symbols is the only one that minimizes the entropy of a description, we refer to it as the Fundamental Scale.

These examples demonstrate how the same message can be symbolically interpreted in several ways. Unavoidably, deciding how to read and to interpret the symbolic information received with the message, is the observer's choice.

*4.5. The Fundamental Scale*

When the communication system works on unknown rules, it is not possible to select *a priori* the set of symbols to interpret the description. That is the case of the texts of any file containing recorded music. There are no words in the sense we are used to and the characters we see do not indicate any meaning for us; we cannot even be sure about the meaning of the space character " ". In natural languages, a space is used as delimiter for words but in music a space does not necessarily mean a silence. Fortunately, even having no idea about the 'grammar' of a communication system, we can rely on the MDL Principle to reveal the symbols with which that communication system is built. Therefore, whenever a description is expressed in a language with unknown grammar, the MDL Principle can be applied in order to unveil the set of symbols which minimize the entropy. That is the strategy followed by the FSA introduced by Febres and Jaffe [9] to obtain measurements of the quantity of information associated with a description based on a totally unknown language.

*4.6. Scale Downgrading*

Once the observer's interpretation settles onto a specific scale, a selection of the relevant symbols forming the description has been established. Accounting for the frequency of each symbol leads to the construction of the symbol probability distribution which characterizes the description. If these probabilities are ordered, a so-called symbol probability profile can be obtained. These profiles are by themselves descriptions of the system observed, represented by a shape of $D-1$ degrees of freedom, where $D$ is the symbolic diversity. When the symbol selection obeys the minimal entropy criterion—that is when the observation is performed according to the Fundamental Scale—the profile shape may serve for identity purposes due to the uniqueness condition of the minimal entropy probability distribution.

In order to use the ranked probability profiles for identification purposes, or to compare systems observed at different scales, it may be convenient to '*simplify*' the frequency profile while preserving its overall shape. This is done by replacing groups of points in the original profile by single points. The selection of point sets to be replaced must be done considering the density of points in each profile segment, in order to keep the same level of detail and significance in all sectors of the profile; this is the basis for the formulation of the scale downgrading process presented in [14].

This procedure for downgrading the language scale is useful given the frequent requirement for expressing descriptions at the same scale or to show any description's profile shape, at any arbitrarily selected scale. Figure 1 shows the profile shape representing the Beethoven's 9th Symphony 3rd Movement, recorded as a MIDI file. Figure 1a shows the frequency profile of the 2828 fundamental symbols which describe the MIDI file at the lowest entropy. Figure 1b,c shows the profile with scale downgraded to 513 and 65 symbols respectively.

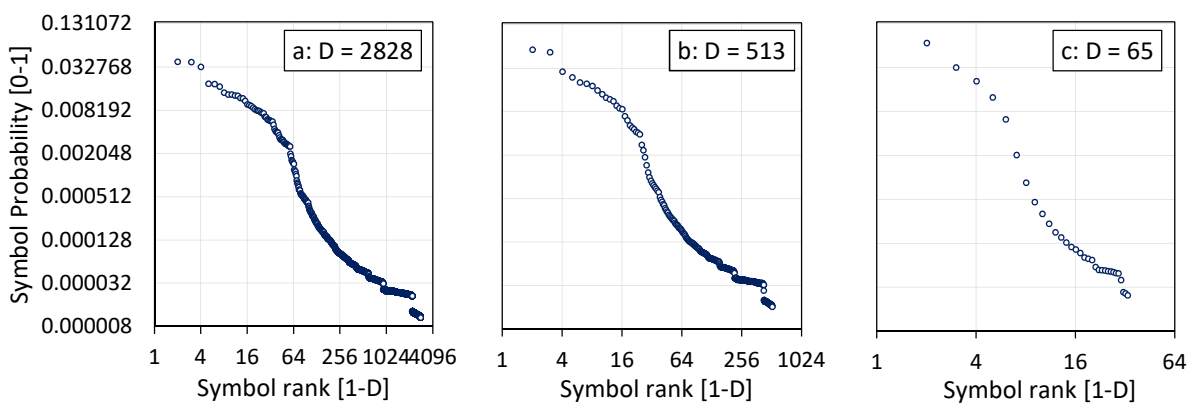

**Figure 1.** The profile representing a Musical Instrument Data Interface (MIDI) version of Beethoven's 9th Symphony 3rd Movement. Figure 1a (**left**) shows the full fundamental scale description having 2828 different symbols; Figure 1b (**middle**) and Figure 1c (**right**) represent the degraded versions of the same profile with 513 and 65 symbols respectively.

## 5. Some Experiments with Different Language Expressions

The following sections present tests where the information of several descriptions is evaluated at different scales of observations. These tests allow for the comparison of the symbolic information resulting from different interpretations of the same descriptions. After recognizing the scope $L$, resolution $R$ and the scale $D$, the corresponding entropy $h$ to each one of these observations is computed by means of Equation (2) and presented in tables to discuss and draw conclusions.

It is worth to emphasize the fact these numerical experiments rely on the application of the concepts of scale and resolution. Prior to calculating the entropy of a description's observation, the number of different symbols—the symbolic diversity—must be known. This number characterizes each observation. Being the entropy a measure of the extractable information from a description, presenting these experiments signals that the concept of scale offers relevant advantages.

*5.1. Natural Languages*

Figure 2 shows the relationship between entropy and message length measured in symbols, for 128 English speeches and 72 Spanish speeches. Details about the text's contents and properties were published by Febres and Jaffe [14].

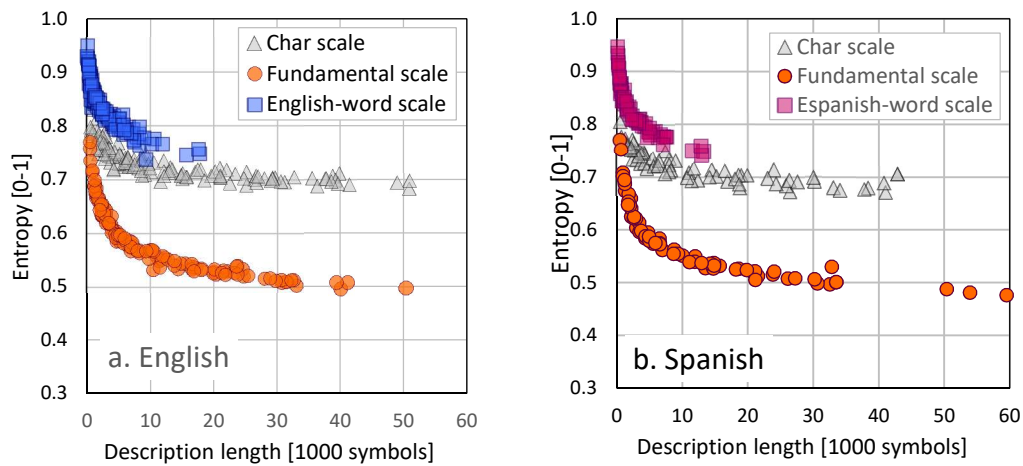

**Figure 2.** Entropy $h$ vs. description length $L$ in symbols. Graphs show the relationship between entropy $h$ and scope $L$ (symbol length) for descriptions expressed in: (**a**) English; (**b**) Spanish. The effect of the observation scale over the resulting observed entropy is shown for three observation scales: characters, words and the fundamental scale.

We know that representing each one of these speeches in a binary scale, will show about the same number of zeroes and ones. As a result, in a binary scale the computed symbolic entropy will be near 1 and therefore, if symbols are not grouped—as the computers do by forming Bytes of seven or eight bits each—it is practically impossible to extract information.

Three other scales were tested: Char, Word and the Fundamental Scale. Among them, the minimal entropy was obtained when reading the message at the Fundamental Scale. Figure 2 illustrates how for any description length, the lowest entropy occurs when the message is observed at the Fundamental Scale. That does not surprise. The symbols at the Fundamental Scale were identified having the minimization of entropy as the objective. It is worth to mention the dramatic difference between the entropy of these observations on the Fundamental Scale and the observations of the same speeches at any other scale.

*5.2. Same Symbolic Structure. Different Perceptions*

Both mosaics shown in Figure 3 are built with identical number of pixels. Each one is an array of 56 × 56 pixels some of which are dark or light colored. For both mosaics, there are white or gray pixels inducing the interpretation of the figures in one or other manner. However, since the number of pixels and different colors used are the same, implying their symbolic information is the same.

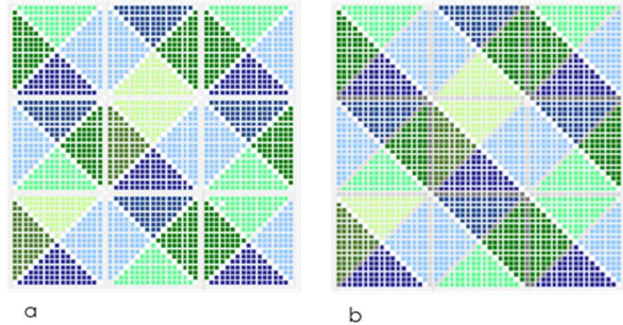

**Figure 3.** Two perceptions of a 2D mosaic with a resolution 56 × 56 pixels. Mosaic (**a**) shows pixels with four different colors. Same color pixels are grouped and separated by white pixels forming triangles; Mosaic (**b**) shows vertical and horizontal white lines dimmed to gray.

When estimating the account for information expressed in each mosaic, however, there are different accounts which depend on the type of information considered. In Figure 3a we interpret the mosaic as a group of 3136 pixels (56 × 56) each one representing one color, out of four possible colors. But changing the focus from single pixels to the larger tiles suggested by the arrangements, triangles in Figure 3a and bands in Figure 3b, the distribution of the types of information settles on the amounts shown in Table 2. Notice that resolution, scope and scale have different values for the interpretations of these mosaics. Yet the entropy equals one for all of them due to the uniform distribution of symbol frequencies; according to Equations (2) and (3) any uniform probability distribution leads to maximum entropy.

**Table 2.** Properties of each interpretation of 2D patterns shown in Figure 3.

| Figure | Figure 3a,b | Figure 3a | Figure 3b |
|---|---|---|---|
| Scale Name | Pixels | Symbols | Symbols |
| Data Representation | Pixels | Triangles | Diagonal Bands |
| Resolution $Rhorz$ | 56 | 3 | 6 |
| Resolution $Rvert$ | 56 | 3 | 1 |
| Resolution $Rangle$ | - [1] | 4 | 1 [2] |
| Resolution $Rcolor$ | 4 | 4 [2] | 2 |
| Scope (Length) $L$ | 3136 | 36 | 6 |
| Scale (Diversity) $D$ | 4 | 4 [3] | 2 [†] |
| Entropy $h$ (0 to 1) | 1 | 1 | 1 |
| Specific diversity $d$ | 0.001 | 0.111 | 0.333 |

[1] This degree of freedom does not exist for single pixels; [2] Only four angular positions are required; [3] Triangles: light blue, light green, dark blue, dark green; [†] Light band, dark band.

*5.3. Partial Changes of Resolution and Scope*

Figure 4 uses a 2D example to illustrate the same picture observed for different combinations of resolution and scope. Figure 4a shows a set of 2D symbols over a 'surface' of 28 × 28 pixels. Here the squared pixels have the role of elementary information and each of them may have one of two values: black or white, then the number of possible states for each square is $D = 2$. There are 28 squares per side, thus the resolution can be approximated to $R = 28 \times 28$ pxl.

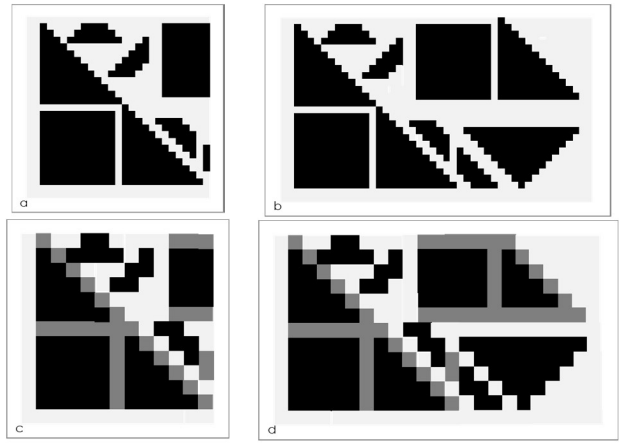

**Figure 4.** Effects of changes of resolution and scope over a 2D representation of polygons. Graphic representation of a language scale downgrading from scale $D$ to scale $S$ ($S < D$). The total number of symbols at scale $D$, representing $D$ different symbols on graphs (a) and (b) (at the top of the Figure), are transformed into $S$ different symbols when the language is represented on the scale $S$, as in the graphs (c) and (d) (at the bottom of the Figure). Also, graphs (b) and (d) (on the right side of the Figure) exhibit greater scope $L$ than graphs (a) and (**c**) (on the left side of the Figure). Keeping the observation distance applied in Figure 4a, Figure 4b increases the scope up to 28 × 46 pixels. Figure 4c reduces the resolution with respect to Figure 4a, d increases the scope and reduces the resolution simultaneously. The impact of these variations of resolution and scope over the entropy and the diversity is presented in Table 3.

**Table 3.** Balance of information for the 2D example presented Figure 4.

| Figure | Figure 4a | | Figure 4b | | Figure 4c | | Figure 4d | |
|---|---|---|---|---|---|---|---|---|
| **Scale Name** | **Pixels** | **Symbols** | **Pixels** | **Symbols** | **Pixels** | **Symbols** | **Pixels** | **Symbols** |
| **Data Representation** | 0's & 1's | Polygons | 0's & 1's | Polygons | 0's & 1's | Polygons | 0's & 1's | Polygons |
| Resolution $R_{horz}$ | 28 | 28 | 46 | 46 | 13 | 13 | 23 | 23 |
| Resolution $R_{vert}$ | 28 | 28 | 28 | 28 | 13 | 13 | 13 | 13 |
| Resolution $R_{angle}$ | -[1] | 8[2] | -[1] | 8[2] | -[1] | 8[2] | - | 8 |
| Scope (Length) $L$ | 784 | 8[2] | 1288 | 10 | 169 | 8 | 299 | 8 |
| Scale (Diversity) $D$ | 2 | 5[3] | 2 | 3[†] | 3 | 4[††] | 3 | 5[†††] |
| Entropy $h$ [0 to 1] | 0.970 | 0.928 | 0.999 | 0.646 | 0.934 | 0.813 | 0.940 | 0.861 |
| Specific diversity d | 0.003 | 0.625 | 0.002 | 0.300 | 0.018 | 0.500 | 0.010 | 0.625 |

[1] This degree of freedom does not exist. The concept of single pixels' angular position degenerates; [2] Only eight angular positions are required to describe symbols represented; [3] Square, triangle, trapezoids, large rectangle, small rectangle; [†] Square, triangle, trapezoids; [††] Large Triangle, rectangle and noise: trapezoids, stairs-like polygon; [†††] Large Triangle, small triangle, wedge and noise: trapezoids, stairs-like polygon.

*5.4. The Impact of Reorganizing*

This section presents a little experiment. The purpose is to assess the impact of organizing the symbols with which we interpret a description. Figure 5 shows arrays of 30 squares colored with five different intensities; the lightest named as 1 and the darkest named as 5. In the leftmost array, Figure 5a, the squares are randomly organized while Figure 5b shows the colored squares ordered with the darkest at the top-left corner of the array and the lightest at the bottom-right corner. Figure 5c,d show the organized distribution of squares indicating different symbols formed by grouping several squares into each type of symbol.

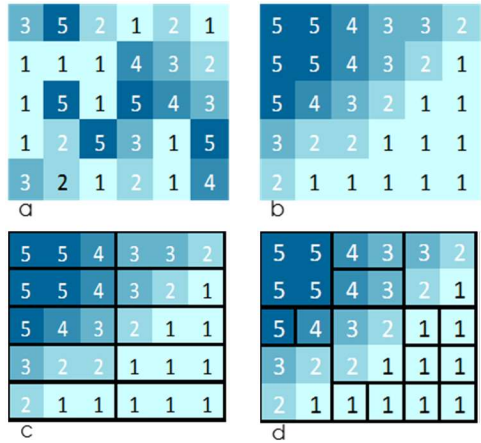

**Figure 5.** Four interpretations of the same distribution of 30 squares colored with five different tones of blue. Numbers indicate each tone used. The lightest is represented by 1 and the darkest with 5. Each tone appears with the same frequencies in the four graphs. (**a**) Shows the 30 squared randomly ordered; (**b**) The squares are ordered according to the rule indicating that no darker square can appear below or at the right of another lighter square; (**c**) Shows groups of symbols formed by a regular shaped lattice of 1 × 6 bricks; (**d**) Shows with black borders the groups of squares forming symbols to reduce the entropy of this interpretation.Table 4 shows the estimates of the entropy assotiated with these four interpretations of the 30 squares .

**Table 4.** Properties of several interpretations of the 2D patterns shown in Figure 5.

| Figure | Figure 5a | Figure 5b | Figure 5c | Figure 5d |
|---|---|---|---|---|
| Scale Name | Symbols | Symbols | Symbols | Symbols |
| Data Representation | Single Squares | Single Squares | Organized Squares | Organized Squares |
| Resolution $R_{horz}$ | 6 | 6 | 2 | 3 |
| Resolution $R_{vert}$ | 5 | 5 | 5 | 3 |
| Resolution $R_{color}$ | 5 [1] | 5 [1] | Varies [2] | Varies [3] |
| Scope (Length) $L$ | 30 [4] | 30 [4] | 10 | 16 |
| Scale (Diversity) $D$ | 5 [1] | 5 [1] | 7 [†] | 6 [††] |
| Entropy $h$ (0 to 1) | 0.943 | 0.943 | 0.970 | 0.812 |
| Symb. Info. $Yh$ [bits] | 0.167 | 0.167 | 0.700 | 0.375 |

[1] Different colors for 1 × 1 array of squares; [2] Approximation of different combinations of ordered 3 × 1 squares; [3] Different combinations of 2 × 2, 2 × 1, 1 × 2 and 1 × 1 arrays of ordered squares; [4] 30 = 1's + 2's + 3's + 4's + 5's = 11 + 6 + 5 + 3 + 5; [†] 5 5 4 | 5 4 3 | 3 3 2 | 3 2 2 | 3 2 1| 2 1 1 | 1 1 1; [††] 5 5 − 5 5 | 4 3 | 3 2 − 2 1 | 5 | 4 | 1.

This test shows the impact of the interpretation over the distribution of the different types of information. Interpreting Figure 5b as an array of 30 squares, requires just as much symbolic information as the disorganized squares presented in Figure 5a. Despite our unavoidable tendency to appreciate order in Figure 5b, if we consider all squares as independent single symbols, transmitting this information would require the same effort as for transmitting Figure 4a. However, if we let our brain group the squares in repeated patterns by degrees of color intensity (see Figure 5c), we reduce the number of symbols we have to consider and the spatial information associated with them. The possibility of associating semantic information also appears along with the variety of different symbols that now can be arranged. This transference of information from one type to another may be augmented (as illustrated in Figure 5d) or diminished with the grouping of the squares to form symbols of any shape.

*5.5. Music*

Table 5 shows a comparison of information balance of two segments of music recorded in .MP3 format. The two segments correspond to the same fraction of Beethoven's 5th Symphony 1st Movement.

The segments differ in the instruments used to play them; the first is played by the full orchestra the piece was written for, while the other is played with a piano solo. As descriptions of these music segments, actual MP3-encoded sound files were used. Figure 6, showing a fraction of one of these files, is included to give an idea of how these files may look like. For each version of the music segment analyzed, three observation scales were used: binary, characters and the fundamental scale. The characters' scale consists of splitting the music-text file in single characters. Each character exhibits a frequency with which entropy is computed. The binary observation can be obtained by substituting each character with its corresponding ASCII number expressed in the binary base. The set of symbols comprising the Fundamental Scale is obtained applying the FSA [9].

**Table 5.** Effects of different observation scales over the quantity of information of a segment of Beethoven's 5th Symphony versioned by a full orchestra and piano solo.

| Beethoven's: | 5th Symphony. 1st Mov. Segment. Orch | | | 5th Symphony. 1st Mov. Segment. Piano | | |
|---|---|---|---|---|---|---|
| Scale Name | Binary | Characters | Fundamental | Binary | Characters | Fundamental |
| Data Representation | Zeroes and Ones | Letters, Punct. and Other Signs | Recognized Min. Entropy Symbols | Zeroes and Ones | Letters, Punct. and Other Signs | Recognized Min. Entropy Symbols |
| Resolutn. $R$ (symb./s) | 188,948 | 23,618 | 4517 | 192,669 | 24,084 | 4241 |
| Scope $LD$ | 5,668,432 | 708,554 | 135,519 | 7,514,080 | 939,260 | 165,387 |
| Scope $L2$ | 5,668,432 | 5,668,432 | 1,084,152 | 7,514,080 | 7,514,080 | 1,323,096 |
| Scale value Diversity $D$ | 2 (0 & 1) | 252 | 4635 | 2 (0 & 1) | 257 | 13,808 |
| Symbolic entropy $h$ | near 1 | 0.990 | 0.893 | near 1 | 0.990 | 0.722 |
| Specific Diversity $d$ | 0.049 | 0.00036 | 0.03420 | 0.049 | 0.00027 | 0.08349 |

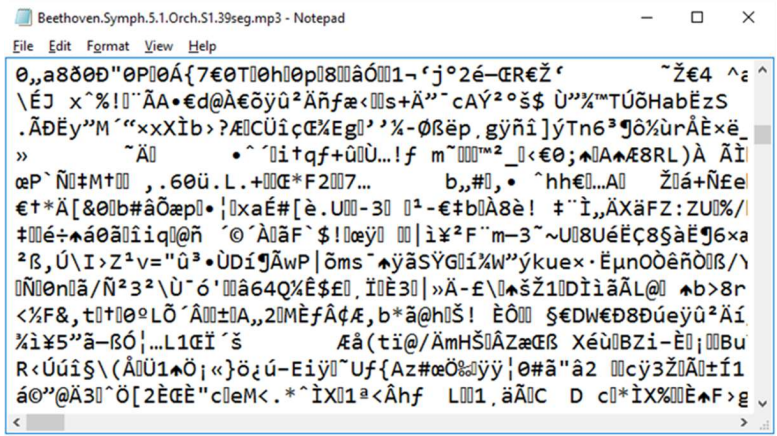

**Figure 6.** A tiny fraction of the text which constitutes the Beethoven's 5th symphony 1st movement segment interpreted with orchestra. Extracted from an .MP3 sound recorded file.

The results show that none of the two pieces analyzed is based on characters. While the entropy for both descriptions, computed with a character-based language, is close to the maximum, indicating no symbolic information associated, a substantial reduction of entropy was reached when descriptions were read at the Fundamental Scale. Reading these texts was possible only using the FSA [9] as a tool to recognize where to focus. Despite not recognizing any meaning of the fundamental symbols encountered, the FSA provides a way of reading these descriptions, so that some structure implying information emerges from initially unreadable texts.

## 6. Discussion

*6.1. Advantages of a Quantitative Notion of Scale*

The term *scale* is commonly used in a qualitative manner. Expressions like "individual scale", "massive scale", "microscopic scale", "astronomical scale" and many other similar ones, are typically used to characterize the type of interpretation that should be given to certain descriptions. However, their utility relies on our subjective criteria to adequately apply those expressions. Subsequently, this rather diffused conception of scale is of little use for the purposes of precisely specifying the conditions of the interpretation of a description.

As an alternative, we propose a quantitative conception of scale. The scale of a system equals the scale of the language used for its description; the scale of the language equals the number of different symbols which constitute the language. Our proposal allowed us to develop the five little experiments presented above. Recognizing the symbols corresponding to each observation criterion was possible thanks to this conception of scale and the hypothesis of the interpretation process. Computing the numerical value of the entropy of each observation was also possible after using the symbolic diversity as the base of the logarithm of Expression (2). This highlights the convenience of considering the scale as the number of different symbols contained in the interpretation of a system's description. Moreover, the experiments' results strongly suggest that a system's description scale is determined, in the first place, by the observer and in a much smaller degree by the system itself. Observing a system at a specified scale implies the use of a certain number of symbols. Hence, the number of different symbols used in a description is linked with our intuitive idea of scale. Therefore, the term scale can be used as a descriptor of the language by specifying the number of different symbols required by the language to describe the modeled system.

Another important concept, with a close relationship to scale, is resolved. Resolution, as a parameter, represents the density of symbols, repeated or not, that participate in a description. The resolution separates the description's space in many smaller spaces-segments. Each space-segment must be occupied by a symbol, thus, even an 'empty space' is a valid and necessary symbol to be considered.

*6.2. The Fundamental Scale as a Language Descriptor*

Modeling descriptions rely on recognizing the symbols making up the description. Thus, when the relevant symbols of a description do not form regular-shaped groups of elementary symbols defined over the resolution grid, this recognition becomes a major barrier to the modeling. The concept of Fundamental Scale, along with the FSA, makes this difficult task possible. By unveiling the most relevant symbols within a description and its symbolic diversity, the description's fundamental scale can be estimated, leading to the possibility of quantitatively modeling any textual description.

The Fundamental Scale reveals the inner structure of the coding of any language. Take music, for example. We cannot recognize sounds from the texts representing music (see Figure 6 for a segment of music represented as text). Thus, we cannot rely on the concept of a word to look for some structure beneath a text file containing the sounds of a musical piece. Analyzing these text files, which we know contain information because their reproductions can be listened, is a painstaking task. Independently of the limited number of notes, rhythms, instruments and any other dynamic effects represented in the music record, the resulting sounds when the piece is played by the performer, are of an enormous, practically unlimited, diversity; music is the result of the superposition of an incredibly large number of components and is full of subtleties which we can hear but which are not present in the music sheet. The full description of the musical phenomena produces files that, when discretized, are so diverse in terms of symbols, that looking for musical word-equivalents, is fruitless. Yet, applying the FSA we are able to recognize the most relevant symbols the coding system used to build a musical description (as the one shown in Figure 6) and their relative frequencies. These are the parameters we need to construct a graphical profile with which music can be represented and identified; quantitatively modeled.

The MDL Principle, being independent of the alphabet used, supports the validity of the concept of Fundamental Scale as a property intimately related to the structure of the language used, despite the alphabet which serves as a mere instrument to build the written version of the message.

*6.3. Usefulness and Applications of the Concept of Scale*

The knowledge obtained when reading a message is part of the information carried by the message. It is an indication about the influence of the message's observation over the quantity of information received. Thus, by means of the concept proposed, it is possible the quantification of the information obtained from an interpretation. The method strives to identify the observation scale in order to maximize the information retrieved. While other quantitative conceptions of scale, as the one used in the Wavelet Transform, are designed to improve the effectiveness and efficiency of engineering problems, such as the compression of information or the processing-time reduction, the scale concept provided here constitutes a useful tool to identify, evaluate and select ways to interpret the description of complex systems.

Practical applications of the interpretation scale rely on the development of algorithms capable of performing extremely computationally-complex tasks as those depicted by Febres and Jaffe [9]. In its current condition, the Fundamental Scale Algorithm, along with the concept of scale, is very useful to study and learn about the structure of information but real-time information-processing is still far away from being a practical option. Nevertheless, the scale concept allows for a better understanding of the structure underlying the description of systems, thus applications for complex system modeling are foreseeable. Systems where the data include extensive and fuzzy information as, for example, social, economic, biological and even political systems, seem adequate situations where a technology based on this concept may result fruitful.

**7. Conclusions**

Independent and mutually exclusive notions of scale, scope and resolution are provided:

**Scale**: The set of different symbols used in a description. The scale can be numerically expressed as the symbolic diversity $D$ of the system's description interpretation.
**Scope**: The total number of symbols used in a description.
**Resolution**: The density of symbols (alphabet-symbols or encoded symbols) used to create the symbols used in a description.

A quantitative conception of scale is introduced. This conception allows for generalizing Shannon's information expression to include transmission systems based on more than two symbols; an expression that may be useful when evaluating the convenience of transmitting information by means of non-binary communication systems. From a mathematical perspective, envisaging the symbolic diversity as the representative value of the observation scale is a convenient result, since it provides an intuitive reason to use the symbolic diversity as the base of the logarithm in a generalized version of Shannon's entropy equation and being consistent with normalized versions [6] of the original expression. Intuitively, saying a system is observed on the scale $D$ is like saying the system is viewed from a distance at which exactly $D$ different objects—or symbols, can be distinguished and from the system's description.

The scale, the scope and the resolution form an interesting basis to describe the dynamics of our interpretation processes when facing complex descriptions. Considering scale as an essential model parameter reinforces the idea about our capacity of thinking of using our ability to build internal languages with useful symbols. The association of the concept of scale with the number of symbols of the language employed by the observer to interpret a description, offers advantages for the effective representation of complex systems. It also serves as an indication of the mechanisms triggered for our understanding of systems. This is a motivating thought since it paves the road to conceive intelligence as the capacity of finding observation scales—or interpretation scales—which lead to a more effective way of understanding nature and its phenomena. Intelligence, or at least part of it, can be regarded as the ability to find the set of symbols which better interprets a message by means of reducing its complexity and thus leading us to a better understanding of our environment.

The concept of scale, along with the method for scale downgrading, is also a promising way to compare different interpretations of the same system, as well as comparing different systems at the same scale. The Fundamental Scale allows for the construction of near minimal entropy profiles, which may serve as systems' identifiers. These tools allow for future experiments which may help in our understanding about the interpretation process. The results provide a framework for building useful models to understand our interpretation process, lighting the mechanisms explaining the formation of information from arbitrarily shaped and meaningless symbols.

**Conflicts of Interest:** The author declares no conflict of interest.